\documentstyle[aps,amsfonts,amssymb,epsfig]{revtex}
\begin{document}
\twocolumn[\hsize\textwidth\columnwidth\hsize\csname @twocolumnfalse\endcsname
\title{Observation of out-of-phase bilayer plasmons in YBa$_2$Cu$_3$O$_{7-\delta }$}
\author{M. Gr\"uninger$^1$, D. van der Marel$^1$, A.A. Tsvetkov$^{1,*}$, and A. Erb$^2$}
\address{$^1$\it Lab.\ of Solid State Physics, MSC, University of Groningen, 
Nijenborgh 4, 9747 AG Groningen, Netherlands\\ 
{\rm $^2$} D\'epartement de Physique de la Mati\`ere Condens\'ee, University of Geneva, 
CH-1211 Geneva 4, Switzerland}
\date{March 16, 1999}
\maketitle
\begin{abstract}
The temperature dependence of the $c$-axis optical conductivity $\sigma(\omega)$ of 
optimally and overdoped YBa$_2$Cu$_3$O$_x$ ($x$=6.93 and 7) is reported in the 
far- (FIR) and mid-infrared (MIR) range. Below T$_c$ we observe a transfer of spectral  
weight from the FIR not only to the condensate at $\omega $=0, but also to a new peak 
in the MIR.\@ This peak is naturally explained as a transverse out-of-phase bilayer 
plasmon by a model for $\sigma(\omega)$ which takes the layered crystal structure into 
account. With decreasing doping the plasmon shifts to lower frequencies and can be 
identified with the surprising and so far not understood FIR feature reported in 
{\it underdoped} bilayer cuprates. 
\end{abstract}
]
\narrowtext
After many years the discussion about the charge dynamics perpendicular to the CuO$_2$ 
layers of the high T$_c$ cuprates is still very controversial. The role attributed to 
interlayer hopping ranges from negligible to being the very origin of high T$_c$ 
superconductivity\cite{cooper}. There is no agreement about the relevant excitations 
nor about the dominant scattering mechanism. The $c$-axis resistivity $\rho _c$ is much 
larger than predicted by band structure calculations. The anisotropy $\rho_{c}/\rho_{ab}$ 
can be as large as $10^5$ and shows a strong temperature dependence, especially in the 
underdoped regime, which has been interpreted as an indication for non-Fermi liquid 
behavior and confinement\cite{anderson}. This strong temperature dependence is due to 
two different regimes with 
d$\rho _c/{\rm dT}\!<\!0$ for ${\rm T}_c\!<\!{\rm T}\!<\!{\rm T}^{\prime}$ and 
d$\rho _c/{\rm dT}\!>\!0$ for ${\rm T}\!>\!{\rm T}^{\prime}$, with a crossover temperature 
T$^{\prime}$ that decreases with increasing doping. There is some agreement as to the 
phenomenology that $\rho_{c}$ is described by a {\it series} of 
resistors\cite{anderson,ong,terasaki}, i.e., that different contributions have to be added, 
and that the sign change in d$\rho _c$/dT is due to the different temperature dependence of 
the competing contributions. Overdoped YBa$_2$Cu$_3$O$_x$ (YBCO) is often regarded as a 
remarkable exception, as $\rho_{c}/\rho_{ab}$ is only about 50, and d$\rho _c/{\rm dT}\!>\!0$ 
for all ${\rm T}\!>\!{\rm T}_c$. It is an important issue whether a sign change in 
d$\rho_c/{\rm dt}$ at low T is really absent or only hidden by T$_c$ being larger than a 
possible T$^{\prime}$, i.e., whether overdoped YBCO follows anisotropic three dimensional 
(3D) or rather 2D behavior.

The $c$-axis optical conductivity $\sigma_1(\omega)$ of YBCO shows several remarkable 
features\cite{homes,schuetzmann,tajima,hauff,bernhard}: 
(1) It's very low value compared to band structure calculations, reflecting the large 
$\rho_c$. (2) A suppression of spectral weight at low frequencies already above T$_c$ 
in underdoped samples referred to as the opening of a 'pseudogap' (which agrees with 
the upturn in $\rho_c$). (3) The appearance of an intriguing broad 'bump' in the FIR at 
low T in underdoped samples. (4) In overdoped YBCO, the spectral weight of the 
superconducting condensate is overestimated from $\sigma_1(\omega)$ as compared to 
microwave techniques\cite{homes98}. 

In this letter we suggest that most of the above mentioned issues can be clarified by 
modelling the cuprates or in particular YBCO as a stack of coupled CuO$_2$ layers
with alternating weaker and stronger links. This multilayer model fits the measured data 
at all doping levels and at all temperatures.\@ A similar model was proposed for the 
superconducting state by van der Marel and Tsvetkov\cite{tsvetkov}. A {\it transverse} 
optical plasmon was predicted. This model has been verified in 
SmLa$_{0.8}$Sr$_{0.2}$CuO$_{4-\delta}$\cite{shibata}. We report the observation of this 
mode in the infrared spectrum of optimally and overdoped YBCO and propose a common origin 
with the above mentioned 'bump' in underdoped YBCO.\@ 

Single crystals of YBa$_2$Cu$_3$O$_x$ were grown using the recently developed BaZrO$_3$ 
crucibles\cite{erb1}, which in contrast to other container materials do not pollute 
the resulting crystals. Crystals grown using this technique exhibit therefore a superior 
purity ($>$ 99.995 at.\ \%)\cite{erb2}. 
The samples had typical dimensions of $2\times0.5$ -- $0.7$ mm$^2$ in the $ac$-plane. 
The O concentration was fixed by annealing according to the calibration of 
Lindemer\cite{lindemer}. An O content of $x$=7 was obtained by annealing for 400 h at 
300$^\circ $C in 100 bar of high purity oxygen. Annealing in flowing oxygen at 
517$^\circ $C for 260 h produced $x$=6.93. Measurements of the ac-susceptibility 
indicate T$_c$=91 K for $x$=6.93 and 87 K for $x$=7. The width of the transitions 
were 0.2 K and 1 K, respectively.
Polarized reflection measurements were carried out on a Fourier transform 
spectrometer between 50 and 3000 cm$^{-1}$ for temperatures between 4 and 300 K.\@ As a 
reference we used an {\it in-situ} evaporated Au film. Above 2000 cm$^{-1}$ the spectra 
are almost T independent. The optical conductivity $\sigma(\omega)$ was calculated via a 
Kramers-Kronig analysis. 

The measured $c$-axis reflectivity and $\sigma_1(\omega)$ derived from it are plotted 
in Fig.\ 1 for 4 and 100K (solid and dashed black lines). Disregarding the phonons, 
$\sigma_1(\omega)$ shows an almost constant value of about 200 $\Omega^{-1}$cm$^{-1}$. 
A Drude-like upturn is only observed at low frequencies in the overdoped case $x$=7. 
Below T$_c$ a sharp reflectivity edge develops at about 300 cm$^{-1}$ (inset of top 
panel), which had been identified as a Josephson plasmon, a collective mode in a stack 
of Josephson coupled 2D superconducting layers. 
The gradual suppression of $\sigma_1(\omega)$ below about 700 cm$^{-1}$ can be attributed 
to the opening of the superconducting gap. The finiteness of $\sigma_1(\omega)$ at 
all frequencies reflects the $d$-wave symmetry of the gap. 
The {\it increase} of $\sigma_1(\omega)$ between 700 and 1300 -- 1500 cm$^{-1}$ from 
100 to 4 K comes as a surprise. The superconducting phase transition obeys case II 
coherence factors for electromagnetic absorption\cite{tinkham}, i.e., only a 
{\it suppression} of $\sigma_1(\omega)$ is expected for frequencies not too close to 0. 
The difference of spectral weight above and below T$_c$ defined as (for T$<$T$_c$):
\begin{equation}
\omega_{\Delta}^2({\rm T},\omega)=8\int_{0^+}^{\omega}
\left[ \sigma_1({\rm 100K},\omega^{\prime})-\sigma_1({\rm T},\omega^{\prime})\right] 
{\rm d}\omega^{\prime}
\label{spwe}
\end{equation}
is expected to rise monotonically with increasing frequency to a constant value for 
frequencies much larger than the gap. It is common practice to determine the spectral 
weight of the superconducting condensate from this constant value. 
However, in YBCO$_{6.99}$ Homes {\it et al.}\cite{homes98} reported a discrepancy 
between $\omega_{\Delta}$ determined from either this optical sum rule 
($2050\pm150$ cm$^{-1}$) or the microwave surface reactance ($1450\pm50$ cm$^{-1}$). 
To account for this difference the existence of a very narrow normal carrier Drude peak 
with a width smaller than the lowest measured frequency was concluded, which contradicts 
again the microwave measurements showing a very small $\sigma_1(\omega)$\cite{bonn}.
Our data clearly indicate a non-monotonic behavior of $\omega_{\Delta}(\omega)$ 
(insets in Fig.\ 1, see also Ref.\ \cite{tajima}) and a spectral weight transfer from 
low frequencies to a new peak above the phonons. This can naturally be explained by the 
following model for $\sigma(\omega)$ which takes into account the layered structure of 
the cuprates.

We devide the unit cell of YBCO into the intra- and inter-bilayer subcells $A$ and $B$. 
Let us imagine, that a time dependent current is induced along the $c$-direction, the 
time derivative of which is $(dJ_c/dt)$. We define $(dV_{j}/dt)$ as the time derivative 
of the voltage between two neighboring CuO$_2$ layers, i.e., across subcell $j$. Our 
multilayer model corresponds to the approximation, that the ratio $(dV_{j}/dt)/(dJ_c/dt)$ 
is provided by a {\em local} linear response function $\rho_j$ corresponding to the complex 
impedance which depends {\em only} on the voltage variations on the neighboring CuO$_2$ 
layers, and not on the voltages on layers further away. Microscopically this corresponds to 
the condition, that in the normal state the mean free path along $c$ must be shorter than 
the distance between the layers, $l_{j}$. In the superconducting state this should be 
supplemented with the same condition for the coherence length along $c$. In this sense, 
the multilayer model reflects the confinement of carriers in the 2D CuO$_2$ layers. 
Let us treat the current as the parameter controlled by applying an external field. Since 
the current between the layers is now uniform and is independent of the subcell index $j$, 
the electric field average over the unit cell is a linear superposition of the voltages 
over all subcells within the unit cell. This effectively corresponds to putting the complex 
impedances $\rho_j$ of subcells {\em in series},  
$ 
\rho(\omega)\!=\!x_A \rho_A(\omega) +  x_B \rho_B(\omega)
$, 
where the $x_{j}\!=\!l_j/l_c$ are the relative volume fractions of the two subcells,
$l_A\!+\!l_B\!=\!l_c$, and $\rho_{j}(\omega)$ are the {\em local} impedance functions within 
subcells $A$ and $B$. This sum for $\rho(\omega)=[\sigma(\omega)+\omega/4\pi{\rm i}]^{-1}$ 
is very different from the case of a homogeneous medium, where different contributions are 
additive in $\sigma(\omega)\! =\!\Sigma \sigma_j(\omega)$, which corresponds to putting the 
various conducting channels of the medium {\em in parallel}. To illustrate this, let us adopt 
the Drude model for the complex interlayer impedance. In parallel conduction the sum of 
e.g.\ two Drude peaks yields 
\begin{equation}
 \frac{4\pi {\rm i}/\omega}{\rho(\omega)} =  
 1 - \frac{\omega_{p,A}^2}{\omega^2+{\rm i}\gamma_A\omega}
 - \frac{\omega_{p,B}^2}{\omega^2+{\rm i}\gamma_B\omega}
\label{2Drude}
\end{equation}
where $\omega_{p,j}$ denotes the plasma frequency, and $\gamma_j$ labels the damping. This 
results in a single plasma resonance at a frequency $\omega_p^2=\omega_{p,A}^2+\omega_{p,B}^2$, 
i.e., only one longitudinal mode (the zero) survives which is shifted with respect to the zeros 
of the individual components. The transverse mode (the pole at $\omega\!=\!0$) is identical. 
Putting two Drude oscillators {\em in series} in the multilayer model, i.e., using 
$\Sigma x_j \rho_j$ has a surprising consequence.
\begin{equation}
 \frac{\rho(\omega)}{4\pi {\rm i}/\omega} = 
 \frac{x_A}{1 - \frac{\omega_{p,A}^2}{\omega^2+{\rm i}\gamma_A\omega}}+
 \frac{x_B}{1 - \frac{\omega_{p,B}^2}{\omega^2+{\rm i}\gamma_B\omega}}
\label{1/rho}
\end{equation}
Now both longitudinal modes (poles of $\rho_j$) are unaffected, and in between a new 
{\it transverse} mode arises. This transverse optical plasmon can be regarded as an out-of-phase 
oscillation of the two individual components. This mode has been predicted in Ref.\ \cite{tsvetkov} 
for the case of a multilayer of Josephson coupled 2D superconducting layers. The existence of 
two longitudinal modes was confirmed experimentally in 
SmLa$_{0.8}$Sr$_{0.2}$CuO$_{4-\delta}$\cite{shibata}. Note that superconductivity is not a 
necessary ingredient, the optical plasmon appears regardless of the damping of the individual 
components. 

In order to apply the model to the measured reflectivity data we have to include the phonons, 
for which a separation into subcells is not generally justfied, e.g.\ for the $c$-axis bending 
mode of the planar O ions, located on the border between subcells $A$ and $B$. Therefore we 
adopt the following model impedance
\begin{equation}
\rho(\omega)\!=\! 
\sum_j
\frac{x_j}{\sigma_j\!+\!\sigma_{ph}\!+\!\sigma_{M}\!+\!\omega/4\pi{\rm i}},\;\;\;\; 
j\in \{A,B\}
\label{1/sig}
\end{equation}
where $x_A\!=\!0.28$, and $x_B\!=\!1-x_A$ for YBCO.\@ Note that this model reduces to the 
conventional expression for a homogeneous medium commonly used for high T$_c$ superconductors 
if we either set $x_A\!=\!0$ or $\sigma_A\!=\!\sigma_B$. The $\sigma_{A,B}(\omega)$ contain 
the purely electronic contributions with eigenfrequency $\omega_0\!=\!0$ within each subcell.
\begin{equation}
4\pi\sigma_j(\omega) = \frac{{\rm i}\omega_{s,j}^2}{\omega}
+\frac{{\rm i}\omega_{n,j}^2}{\omega+{\rm i}\gamma_j},\;\;\;\; j\in \{A,B\}
\label{sigj}
\end{equation}
where $\omega_{s,j}$ and $\omega_{n,j}$ label the plasma frequencies of superconducting and 
normal carriers, respectively. All other contributions (phonons, MIR oscillators, etc.) are 
assumed to be identical in the two subcells and are included in a sum of Lorentz oscillators.
\begin{equation}
\frac{4\pi {\rm i}}{\omega } [\sigma_{ph}+\sigma_{M}]= 
\sum\frac{\omega_{p,j}^2}{\omega_{0,j}^2-\omega^2-{\rm i}\gamma_j\omega}
\label{sigh}
\end{equation} 
where $\omega_{0,j}$ denotes the $j$-th peak frequency. 

The agreement between the measured reflectivity data and fits using this model is very good 
at all temperatures (thick gray lines in Fig.\ 1). The strong MIR peak of the optical plasmon 
caused by the out-of-phase oscillation of the superconducting carriers in the two subcells 
is very well reproduced. Note that in a conventional Lorentz model the optical plasmon would 
have to be fit with three parameters $\omega_0$, $\omega_p$ and $\gamma$. Also our model has 
three new parameters, namely the two sets of $\omega_s$, $\omega_n$ and $\gamma$ of 
Eq.\ \ref{sigj} for the two subcells as compared to the single set used within a conventional 
two-fluid fit. In the case of $x$=6.93 at 4 K we have $\omega_{n,A}\!=\!\omega_{n,B}\!=\!0$, 
leaving only one new parameter $\omega_s$. 
In Fig.\ 2 we plot the real part of the dynamical resistivity $\rho(\omega)$. The thick gray line 
was obtained from the full fit parameters and agrees with the Kramers-Kronig result. The solid 
line depicts the electronic contribution $\rho_e(\omega)$, which was obtained by leaving out the 
phonon part $\sigma_{ph}(\omega)$ from the fit parameters in Eq.\ \ref{1/sig}. In the multilayer 
model $\rho_e(\omega)$ is the sum of the subcell contributions 
$x_j \rho_{ej}\!=x_j/(\sigma_j+\sigma_{M}+\omega/4\pi{\rm i})$ ($j\in \{A,B\}$, dashed lines), 
which shows that the two peaks in $\rho_e(\omega)$ can be attributed to the plasmon peaks in the 
two subcells. 
Contrary to the conventional model, the different contributions are not strictly additive 
in $\sigma_1(\omega)$ due to the inverse summation in Eq.\ \ref{1/sig}. 
Nevertheless we can calculate an estimate of the electronic contribution $\sigma_e(\omega)$ 
from the fit parameters in the same way as done for $\rho_e$. An estimate of only the normal 
electronic contribution $\sigma_{en}(\omega)$ is obtained by leaving out the London terms 
$\propto \omega_{s,j}^2$ together with $\sigma_{ph}$. The contribution arising from 
the presence of superconducting carriers is then defined as 
$\sigma_{es}(\omega)\!=\!\sigma_e(\omega)\!-\!\sigma_{en}(\omega)$ (see Fig.\ 1). 

With decreasing doping level the absolute value of $\sigma_1(\omega)$ decreases and therefore 
the optical plasmon peak becomes sharper. At the same time, all plasma frequencies and hence 
also the optical plasma mode shift to lower frequencies. 
This scenario explains the strong FIR 'bump' reported in underdoped YBCO\cite{homes,schuetzmann}.
Similar bumps have been observed in other bilayer cuprates\cite{reedyk,basov}, but never in a 
single layer material. This bump has hindered an unambiguous separation of electronic and phononic 
contributions to $\sigma_1(\omega)$. In Fig.\ 3 we show reflectivity spectra of underdoped samples 
of YBCO taken from Refs.\ \cite{homes,schuetzmann} together with fits using the multilayer model. 
Again good agreement with the model is obtained. The strong phonon asymmetries present in the 
underdoped samples called for a fine tuning of the model: 
the two apical O stretching phonon modes at about 600 cm$^{-1}$ were described by {\it local} 
oscillators in the inter-bilayer subcell $B$, i.e., they moved in Eq.\ \ref{1/sig} from 
$\sigma_{ph}(\omega)$ to $\sigma_{B}(\omega)$. The figure demonstrates that this reproduces 
the asymmetry of the experimental phonon line shape well, although a Lorentz oscillator was 
used. Similar fine tuning has only a minor effect on the quality of the fit for the data presented 
in Fig.\ 1. Comparing the various doping levels shows that both the bending (350 cm$^{-1}$) and the 
stretching (600 cm$^{-1}$) phonon modes show strong assymetries whenever they overlap with the 
transverse plasma mode, but that both modes are symmetric if the transverse plasmon is far enough 
away, as e.g.\ in the case of $x$=7. 
Previously it was argued that the phonon spectral weight is only conserved for 
different T if the bump is interpreted as a phonon\cite{homes}. However, a sum rule exists only 
for the total $\sigma_1(\omega)$, not for the phonon part separately. Moreover, in this scenario 
the width of the bump, it's temperature and doping dependence and the phonon asymmetries remained 
unexplained. 

Both the low frequency Josephson plasmon and the bump are suppressed simultaneously by Zn 
substitution\cite{hauff}, which supports our assignment that both peaks are plasma modes. 
An increase of spectral weight of the bump with decreasing T was reported to start 
far above T$_c$\cite{homes,schuetzmann}, but a distinct peak is only observed below T$_c$. 
We obtained good fits for all T (not shown). As mentioned above, superconductivity is not a 
necessary ingredient of the multilayer model, an out-of-phase motion of normal carriers will 
give rise to a peak at finite frequencies, too. Upon cooling below T$_c$, the reduction of the 
underlying electronic conductivity due to the opening of a gap and the reduced damping produce 
a distinct peak.                                   

Our results imply that even the $c$-axis transport between the two layers of a bilayer is 
incoherent, which agrees with the absence of a bilayer bonding-antibonding (BA) transition in 
our spectra. Using photo electron spectroscopy\cite{schabel} a BA splitting of about
3000 cm$^{-1}$ was reported. The anomalous broad photoemission lineshape may explain 
the absence thereof in the optical data.

In conclusion, we observed the out-of-phase bilayer plasmon predicted by the multilayer model.
The good agreement of the optical data with the multilayer model at all temperatures and doping 
levels shows that YBCO can be modelled by local electrodynamics along the $c$-axis in both the 
normal and the superconducting state. This applies even to overdoped YBCO, one of the least 
anisotropic cuprates. Our results strongly point towards a non-Fermi liquid picture and confinement 
of carriers to single CuO$_2$ layers. 

We gratefully acknowledge C. Bernhard and S. Tajima for helpful discussions. 
The project is supported by the Netherlands Foundation for Fundamental Research 
on Matter (FOM) with financial aid from the Nederlandse Organisatie voor 
Wetenschappelijk Onderzoek (NWO). 

$^{*}$ Also P.N. Lebedev Physical Institute, Moscow.

\newpage
\onecolumn

\begin{figure}[t]
\centerline{\psfig{figure=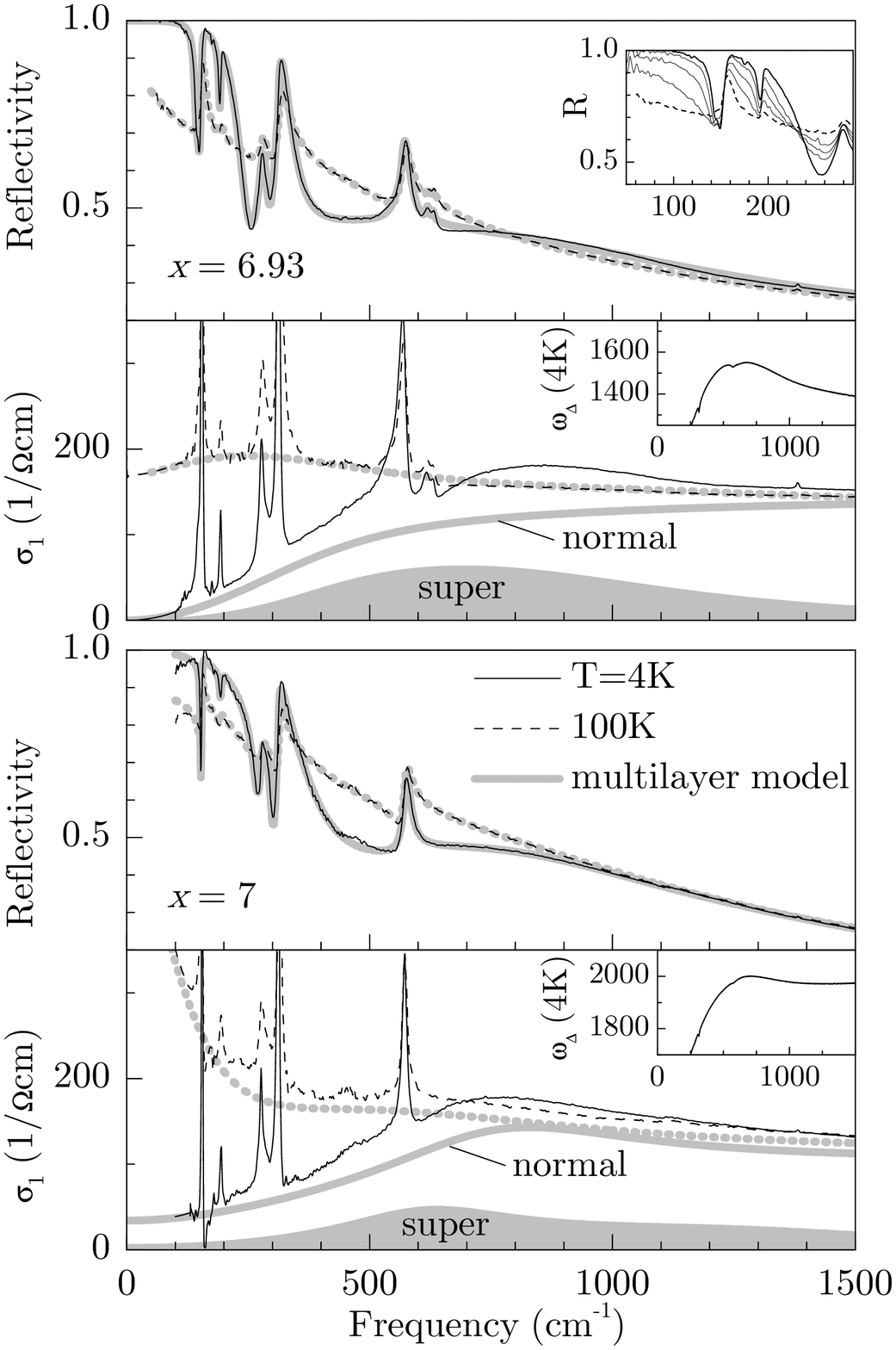,width=13cm,clip=}}
  \caption{The $c$-axis reflectivity $R$ and $\sigma_1(\omega)$ above (dashed lines) 
  and below T$_c$ (solid lines). The thick gray lines depict fits of $R$ using the 
  multilayer model and, in $\sigma_1(\omega)$, the normal carrier electronic contribution 
  $\sigma_{en}(\omega)$ derived from it. The filled areas show $\sigma_{es}(\omega)$ as 
  defined in the text. 
  Insets: $\omega_{\Delta}$(4K,$\omega$) as defined in Eq.\ 1 and detailed 
  T-dependence of $R$. }
\end{figure}

\begin{figure}[t]
\centerline{\psfig{figure=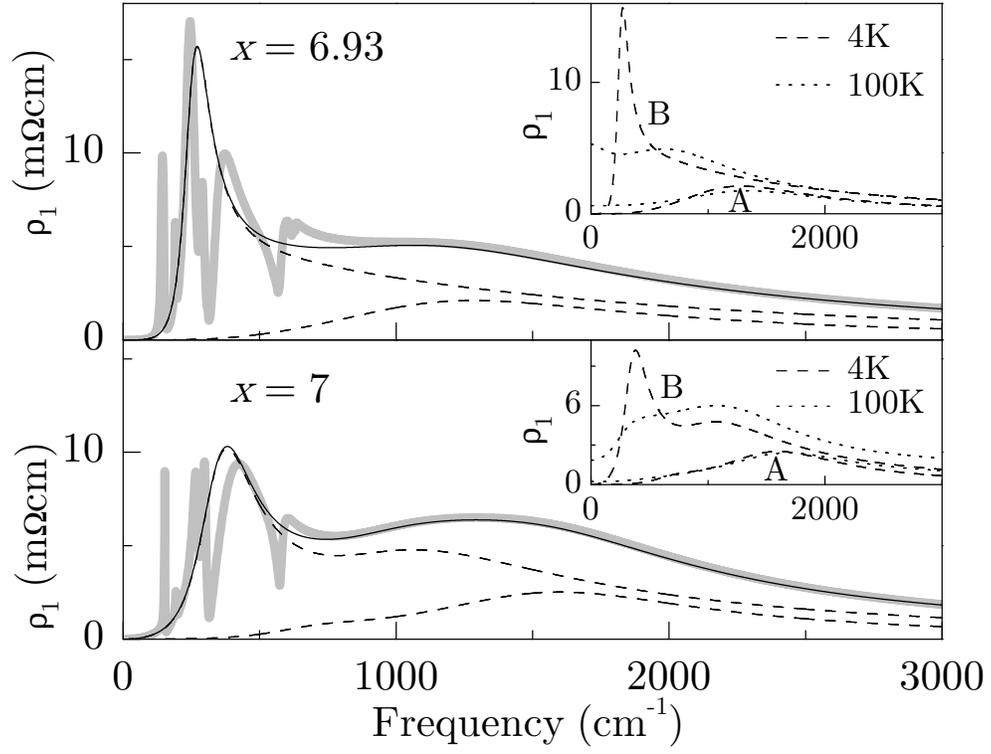,width=13cm,clip=}}
  \caption{Real part of the dynamical resistivity $\rho(\omega)$ as obtained from 
  the multilayer model (thick gray line). The solid line depicts the electronic contribution 
  only, and the dashed lines the subcell contributions $\rho_{eA}(\omega)$ and 
  $\rho_{eB}(\omega)$ to it. Insets: T-dependence of $\rho_{eA}(\omega)$ and 
  $\rho_{eB}(\omega)$. }
\end{figure}

\begin{figure}[t]
\centerline{\psfig{figure=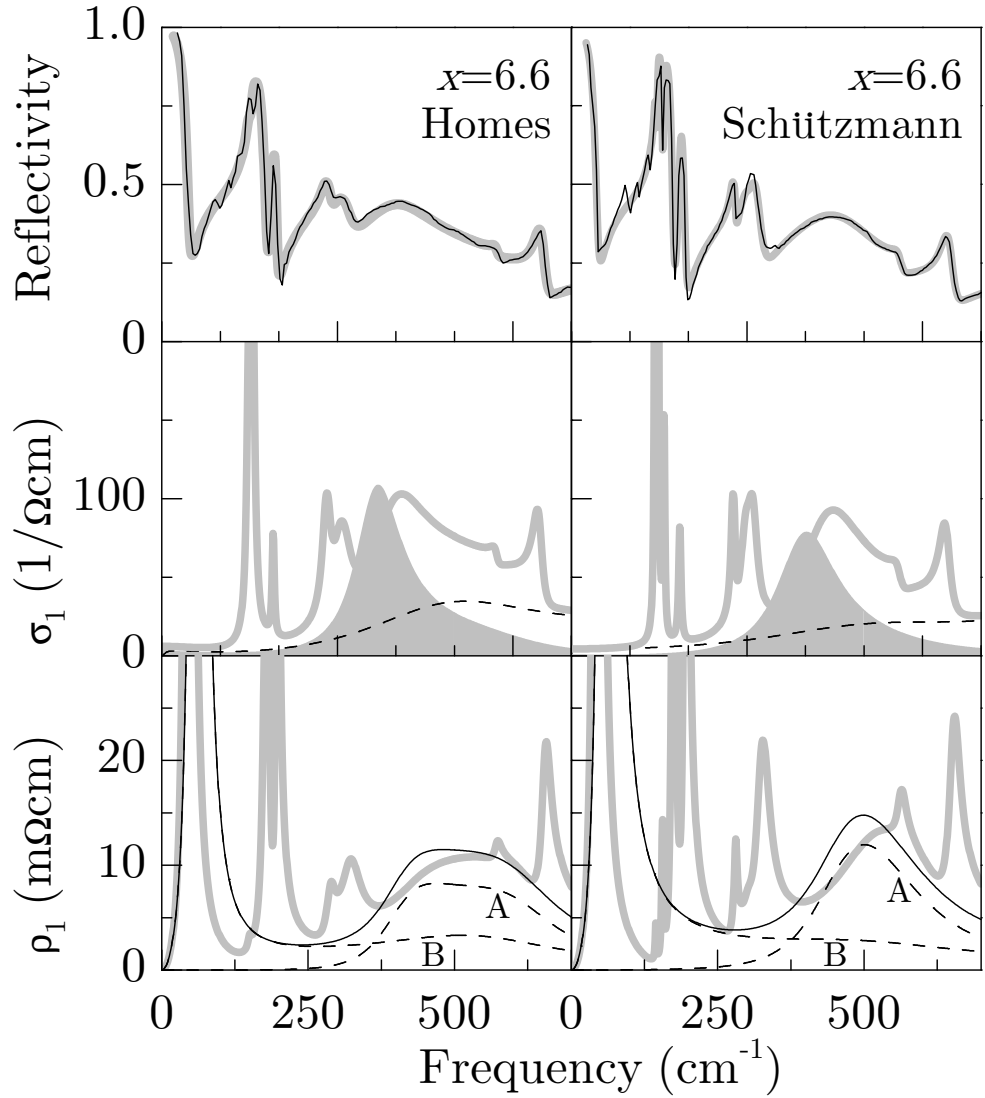,width=13cm,clip=}}
  \caption{Reflectivity data as taken from Refs.\ [5,6] and 
  fits using the multilayer model. In all panels the thick gray lines show the fit result. 
  The mid panels show \protect{$\sigma_1(\omega)$} and the different electronic contributions 
  to it ($\sigma_{es}(\omega)$: filled area, $\sigma_{en}(\omega)$: dashed line). 
  The solid lines in the bottom panels show the electronic contributions to $\rho(\omega)$ 
  and the separation into subcells $A$ and $B$ (dashed lines). }
\end{figure}

\end{document}